\documentstyle[multicol,aps,epsfig]{revtex}
\begin{document}
%\draft
\title{Crosslinked polymer chains with excluded volume: A
new class of branched polymers?}
\author{T.A. Vilgis}
\address{Max-Planck-Institut f\"ur Polymerforschung,
Postfach 3148, 55021 Mainz, Germany}
\date{\today}
\maketitle

\begin{abstract}
In this note microgels with and without
excluded volume interactions are considered. Based on earlier
exact computations on Gaussian mircogels, which are 
formed by self-crosslinking (with $M$ crosslinks) 
of polymer chains with chain
length $N$ Flory type
approximations are used to get first insight to
their behavior in solution. It is shown that two
different types of microgels exist: 
A special type of branched polymer
whose size scales as $R \propto N^{2/5} M^{-1/5}$, instead of
$R \propto N^{1/2}$. The second type are $c^*$ - microgels
whose average mesh sizes $r$ are swollen and form  self avoiding walks
with a scaling law of the form
$r = a (N/M)^{3/5}$.

\end{abstract}

\begin{multicols}{2}

\section{Introduction}

In  recent publications we have shown that the problem
of non-interacting self crosslinked chains can be solved exactly
for a given crosslink configuration \cite{solf1,solf2,solf3}. 
In these papers the size of the crosslinked chain has been calculated
in different regimes of the crosslink coupling strength.
The
results presented there agree perfectly with simulations
carried out independently by Kantor and Kardar \cite{kankar}.
Moreover the method which was employed could be transformed 
to dynamical problems. There exact results can be 
produced also \cite{solf4}.

Bryngelson and Thirumalai \cite{brythi}
had investigated a similar
model of an isolated polymer chain consisting of $N$
monomers subject to $M$ crosslinking constraints by a variational
technique.  It was claimed that a sufficient number of crosslinks
forces the polymer to undergo a transition from an extended ($R^2\propto
N$) to a collapsed state with mean squared end-to-end distance $R^2$ of
order unity. These authors had drawn their conclusion from a
variational computation, but their results to not fully agree
with our exact evaluations \cite{solf1,solf2,solf3}

On the other hand, whenever excluded volume
effects come into play, the problem becomes extremely
difficult and the size of the microgel cannot be computed exactly.
Although the exact "bare" propagator for these networks
is known exactly from the previous work, renormalization group
calculations cannot be carried out easily, because the mathematics
becomes very involved and to carry out systematic perturbation 
expansion seems to be impossible. Apart from the results on the size of
microgels with excluded volume, it is suggested that a gel linked with
weak crosslinks \cite{solf2} defines a new class of branched polymers. These
weak (or soft) crosslinks can be visualized by polymer chains of different
chain flexibility themselves. Although these gels scale in the Gaussian
limit with the same exponent as 
branched polymers  the molecules behave
completely different when excluded volume is present: They do not swell the
same way as classical branched polymers \cite{branched}.

The paper is organized as follows. In the next section we briefly recall the
results on  the
ideal network, which are then reproduced by simple scaling estimates. In
the following we discuss excluded volume effects on the basis of Flory
arguments \cite{pg}. It is shown there 
that the cases of weak and strong crosslinks yield
very different results, in agreement with the physical pictures on branched
polymers and gels.

\section{The ideal network}

We adopt the minimal model of Deam and Edwards \cite{deam,edvi}, i.e.,
a Gaussian polymer chain that is
$M$ times crosslinked to itself (see figure (1)). 
In the Hamiltonian only
terms that model chain connectivity and contributions due
to crosslinking are retained.  Complicating factors such as
entanglements, excluded volume are deliberately neglected
from the start. An appropriate discrete Hamiltonian to begin with is
\begin{equation}
\label{1}
\beta {\cal H}_0=\frac d{2a^2}\sum_{i=1}^N({\bf R}_i-{\bf
R}_{i-1})^2+\frac {d}{2\varepsilon ^2}\sum_{e=1}^M({\bf
R}_{i_e}-{\bf R}_{j_e})^2~.
\end{equation}
We have assumed $N+1$ monomers whose locations in space are
given by $d$-dimensional vectors ${\bf R}_i$
($i=0,1,...,N$).  Distance constraints exist between pairs
of monomers labeled by $i_e$ and $j_e$. For further use we
introduce the inverse strength of the crosslinking
potential
\begin{equation}
\label{2}
z=\left(\frac{\varepsilon}{a}\right )^2
\end{equation}
as the mean squared distance between monomers that form the
crosslinks measured in units of the persistence length $a$
of the chain (figure (\ref{lersch})).  
Limiting cases are given by
$z=0$ (hard ${\delta}$-constraints) and $z\rightarrow
\infty$ (free chain). The whole crosslinking topology is
specified by a set of $2M$ integers
C=$\{i_e,j_e\}_{e=1}^M$. It has been shown
\cite{solf1,solf2,solf3} that the model 
in (\ref{1}) is equivalent to the
Deam-Edwards model \cite{deam} without excluded-volume
interaction if averages are understood in the following
sense
\begin{equation}
\label{3}
\Big\langle ....\Big\rangle_0= \lim_{z \rightarrow 0}\frac{
{\displaystyle\int} \prod_{i=0}^N d{\bf R}_i\,e^{-\beta
{\cal H}_0}....}{{\displaystyle \int} \prod_{i=0}^Nd{\bf
R}_i\,e^{-\beta {\cal H}_0}}~.
\end{equation}

\begin{center}
\begin{minipage}{5cm}
\label{lersch}
\centerline{{\epsfig{file=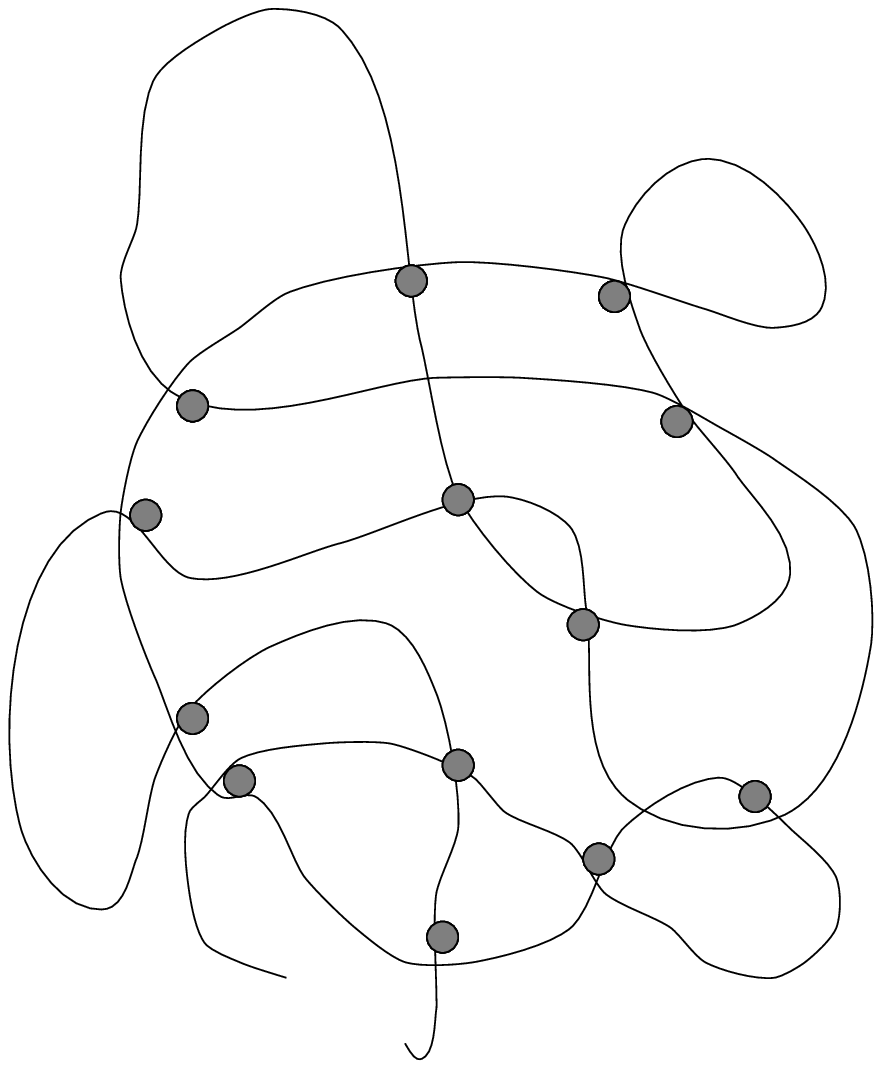,width=4.5cm}}} 
\centerline{{ Fig.(1)
A simple representation of a  microgel}}
\end{minipage}
\end{center}
\vspace{1cm}
To model $M$ {\it uncorrelated} crosslinks the distribution
of frozen variables C is assumed to be uniform
\begin{equation}
\label{4}
{\cal P}(\rm C) =  \prod_{e=1}^M \bigg\{ \frac{2}{N^2}
\sum_{0\leq i_e<j_e\leq N}
\bigg \}
\end{equation}
Other distributions are in principle possible but not
considered in this investigation. As usual for systems with
permanent constraints care must be taken in evaluating
averages of physical quantities. The strategy here is not
to start with the quenched average over the frozen
variables by employing for instance the replica trick, but
to keep explicitly all crosslink coordinates C during the
calculation.  Only at the very end the physical observable
of interest is evaluated for a particular realization of C
which is generated by the distribution in (\ref{4}).
Clearly both approaches will give the same results if only
self-averaging quantities are considered.

The Hamiltonian in Eq. (\ref{1}) together with the uniform
distribution of crosslinks (\ref{4}) defines our working model
for the ideal network.

\section{Scaling Estimates of the Bare Network Hamiltonian}

In this section we recall briefly the exact results 
for the size of the crosslinked
chain. In fact, these can be estimated (apart from
uninteresting constant) from the Hamiltonian (\ref{1}) by simple Flory
type arguments. We are here in the lucky position that the scaling results can
be checked directly with the corresponding exact computations and will provide
additional confidence, when we turn to the excluded volume case shortly below.

\begin{center}
\begin{minipage}{4cm}
\label{soft}
\centerline{{\epsfig{file=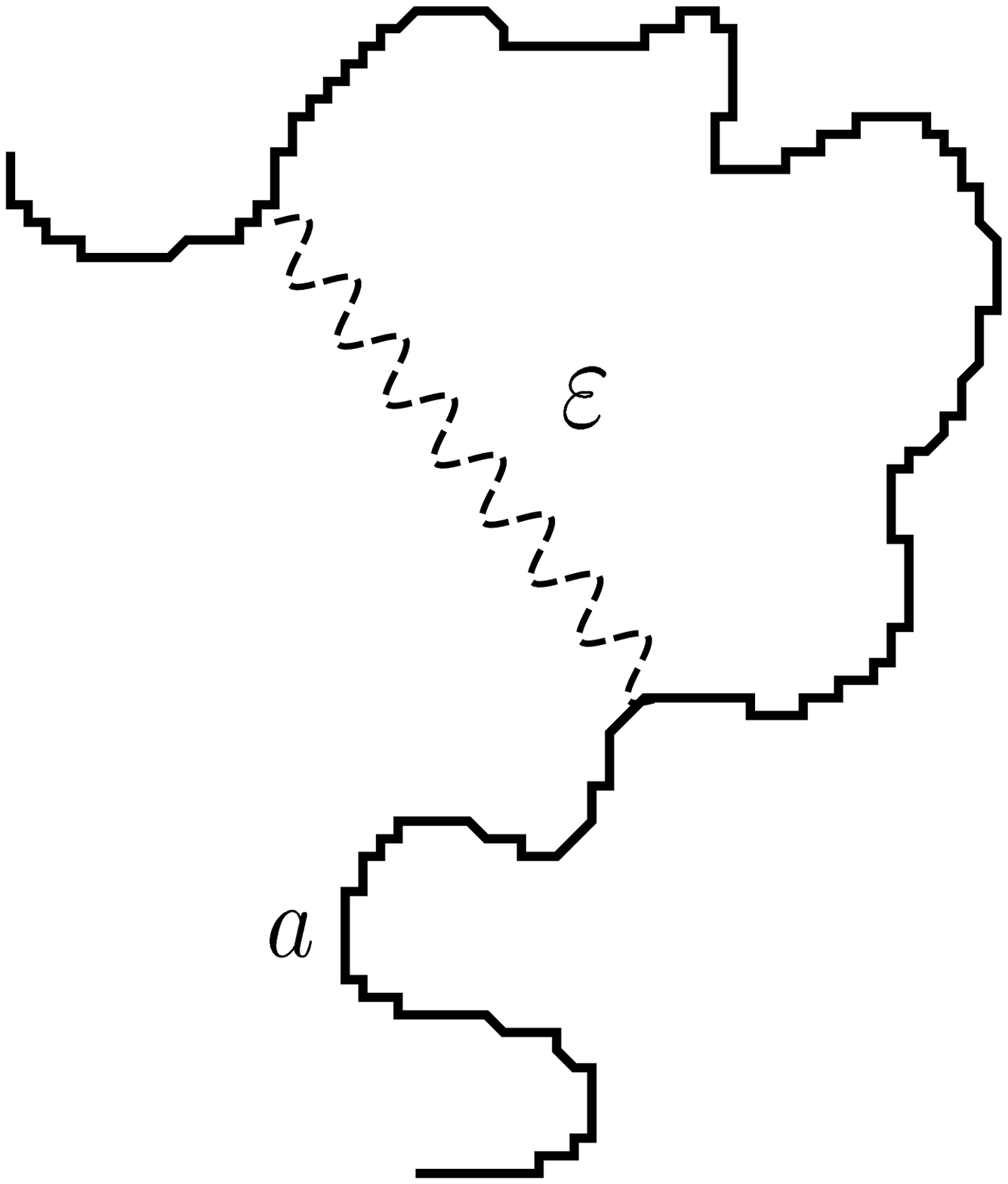,width=3.5cm}}} 
\centerline{{ Fig.(2): The role of $\varepsilon$.\\
}}
\end{minipage}
\end{center}
\vspace{1cm}

These results will provide 
a scaling basis for considerations for gels with excluded volume.
Here we claim that
the Hamiltonian given by Eq.(\ref{1}) contains three different regimes.
These regimes appear naturally: The first will be the trivial free chain
limit, that appears naturally if the crosslink constraint disappears for 
$\varepsilon \to \infty$. A second regime can be expected, when the crosslinks are
soft, i.e., corresponding to a finite and relatively large value of
$\varepsilon$. The microgel regime itself will correspond to the case of hard
crosslinks $\varepsilon \to 0$. Let us now going to 
discuss these three cases in
more detail. 
\begin{enumerate}
\item{free polymer regime: $\varepsilon \to \infty$}\\
\\
This regime is the trivial solution of the problem
and corresponds to vanishing or alternatively to
very weak crosslink constraints. Thus
the free chain result
$$
R_{\rm g}^2 = \frac{a^2}{6}N
$$
must be recovered.
\item{branched polymer regime: $\varepsilon \sim {\cal O}(N^{1/2})$}\\
\\
The connectivity term models
the standard entropic elasticity of a Gaussian chain, i.e.,
$R^2/(Na^2)+N a^2/R^2$, where $R$ is a measure of the size
of the system.  The first term accounts for  stretching,
whereas the second term describes the response due to
compression \cite{pg}.
For relatively soft crosslinks, i.e. $\varepsilon >> a$,  
the crosslink term  requires appropriate attention. 
In the regime of relatively large $\varepsilon$ the second term of the
Hamiltonian is estimated by simple dimensional
analysis to $M (R/\varepsilon )^2$,
because the mean squared distance between a pair of constrained
monomers is of order $\varepsilon^2$. Thus, the physically 
relevant parts of
the total Flory free energy are then
$$
{\cal F} \approx \frac{a^2 N}{R^2}
+
M \frac{R^2}{\varepsilon^2}~.
$$
The first term corresponds to the entropy penalty
due to the chain  shrinking of the effectively "attractive" crosslinks.
That means we consider the only values for $\varepsilon$ in the limits
\begin{equation}
\label{limi}
\sqrt{N} < \frac{\varepsilon}{a} < \sqrt{NM}
\end{equation}
The second term of the 
free energy describes the harmonic potential fro the crosslink constraint. 
Minimization yields the at first sight surprising
result
\begin{eqnarray}
\label{bra}
R_{\rm g} \cong a  \sqrt{ \frac{\varepsilon}{a}}
\left( \frac{N}{M} \right)^{1/4}~.
\end{eqnarray}

This result is in a way surprising in the sense that the
same scaling exponent appears like in the ideal branched
polymer. This is indeed physically sensible, since the
soft crosslink potential connects loosely $M$ monomers
and changes the connectivity of the originally
linear chain to a randomly branched polymer object.

On the other hand the branched polymer result suggested the mapping
on the classical branched polymer limit introduced by Stockmayer and Zimm
\cite{stocky},
although in the computation of these authors no loops are present. In contrast
to the mircogels build up of self crosslinked chains, where many loops form
the network. Nevertheless it is tempting to assume that the branching index
$\Lambda$ defined by Stockmayers size of the ideal branched molecule via
$R \cong a (N/\Lambda)^{1/4}$. Comparison with eq.(\ref{bra}) suggests 
$\Lambda \cong M(a/\varepsilon)^{2}$. It is interesting to note that
the two natural limits on $\Lambda$, i.e., those from the classical 
theory of branched polymers 
$$
\Lambda =\cases{
1/N & linear chain \cr
M/N & branched limit
}
$$
reproduce exactly the limits on $\varepsilon$ in eq.(\ref{limi}). This 
surprising fact shows that for the configurational scaling properties on the
overall size of the gel loops do
not seem to play a significant role. For the latter statements 
we had of course we have ignored here the
detailed internal structure of the microgel. Instead we had considered only
the overall size of the crosslinked chain.

\item{network regime: $\varepsilon \to 0$}\\
\\
The case of hard crosslinks $\varepsilon \simeq {\cal
O}(a)$ is more difficult to obtain.  
To find a reasonable estimate we picture the system
as a coarse-grained random walk over the $M$ crosslinks
with an effective step length proportional to $N/M$, i.e.,
the mean number of monomers between crosslinks. Then 
the crosslink term is estimated to be
of the order $M [R^2/(a^2 N/M)]$. The latter expression has
the effect that it tries to shrink the chain upon cost of
confinement entropy. A suitable Flory free energy is given
by
$$
{\cal F} \approx \frac{a^2 N}{R^2}
+
M \frac{R^2}{a^2(N/M)}
$$
and the size of the network is given by
\begin{eqnarray}
R_{\rm g} \cong a  
\left( \frac{N}{M} \right)^{1/2}
\end{eqnarray}
\end{enumerate}

We have performed a more extensive and exact study on the same working
model \cite{solf1,solf2,solf3}. These scaling estimates agree with the exact
computation and are summarized for completeness:

Based on an exact theorem derived in Ref. \cite{solf1,solf2} 
we have calculated
the radius of gyration $R_{\mbox{ g}}$ of a chain of $N$ monomers
with $M$ uncorrelated crosslinks. Our derivation also allowed for
variation of the strength of the constraint $z=(\varepsilon /a)^2$, where
$a$ denotes the persistence length of the chain and $\varepsilon$ the mean
distance between two monomers that form a crosslink.  Thus, in our model
$z=0$ corresponds to delta constraints (the case usually considered in the
literature), $z=1$ yields soft constraints, and $z\rightarrow \infty$
leads to the ideal chain.  From this study we could clearly distinguish
between three different scaling regimes and these are nicely in agreement with
the naive scaling estimates.
\begin{equation}
R_{\rm g}^2/a^2 \propto
\left \{ \begin{array}{ll}
N/M~~~\mbox{(hard gel)} & \mbox{,~~~if}~~z\lesssim 1\\
z \sqrt{N/M}~~~\mbox{(branched)} & \mbox{,~~~if}~~1 << z << M\\
N~~~\mbox{(free chain)} & \mbox{,~~~if}~~z\gtrsim M~.\\
\end{array} \right .
\end{equation}
In particular, our investigation showed that the cases $z=0$ (hard
constraint) and $z=1$  only differ by a numerical
prefactor which varies from 0.26 ($z=0$) to 0.27 ($z=1$).  Thus
$R_{\rm g}^2/a^2\propto N/M$ for the network situation which
seems to be at variance with the
conclusions by Bryngelson et al. \cite{brythi} who argued that a
critical number of crosslinks $M\geq M_c\sim N/\log N$ will cause the
polymer chain to collapse \cite{brythi}. Our result is in agreement with
recent MC simulations by Kantor and Kardar \cite{kankar} who found for the
end-to-end distance $R^2=1.5N/M$.  This indicates the same one to six
ratio $R_{\mbox{ g}}^2/R^2$ as for ideal polymers without crosslinks
and excluded-volume interaction. For completeness, for $z\rightarrow \infty$ we
recover the case of a free chain with $R_{\mbox{ g}}^2/a^2=N/6$.

The polymer subject to internal crosslinking constraints is
collapsed in a sense that $R_{\mbox{ g}}$ is always proportional to
the square root of $N/M$. This corresponds somehow also to the random
crosslinking process, especially by the random choice of the two monomer
pairs, which form a crosslink.
In the following we are going to discuss the effect of excluded volume. 
If the ideal limit is left and self avoidance are taken into
account, firstly, the appearance of the ratio $N/M$ is no longer obvious and
secondly the space dimension and the number of crosslinks will play a crucial
role.

\section{Excluded volume effects}

The Hamiltonian of the network is must e extended by the total repulsive
excluded volume of all monomers and in given by
\begin{eqnarray}
{\cal H}& =& {d\over 2a^2}\sum_{i}^N\left( {\bf R}_i - {\bf R}_{i+1}\right)^2
\nonumber \\
          & +&{d\over 2\varepsilon^2}
           \sum_{e=1}^M\left({\bf R}_{i_e} - {\bf R}_{j_e}\right)^2\\
          & +&{v\over 2} \sum_{i\neq j}^N 
\delta\left( {\bf R}_i - {\bf R}_{j}\right)  \nonumber                   
\end{eqnarray}
The first two terms correspond to the non interacting network as above
and the last term in the Hamiltonian is the total excluded volume energy.
Although the exact propagator of the bare problem is known,
there is no realistic
hope that excluded volume effects can be treated in a sophisticated
way. In fact, the bare propagator of the problem can only be represented in a
matrix form, and is not easy to handle for analytical purposes. 
Thus we only employ Flory estimates in the following section to get a first
insight about the effects of the excluded volume interactions.

\subsection{Soft crosslinks $\varepsilon >> a$}

To do this we estimate the terms from the above equation by the free energy
\begin{equation}
{\cal F} \simeq M \frac{R^2}{\varepsilon^2} +
v \frac{N^2}{R^d}
\end{equation}
where $d$ is the space dimension.
The latter term in the free energy is the well known excluded volume term
for a chain with $N$ segments. 
Minimization of the free energy yields the size of the microgel
\begin{equation}
R \cong a \left( \frac{\varepsilon}{a} \right)^{1/(d+2)}
\left( \frac{N^2}{M}\right)^{1/(d+2)}
\end{equation} 
or in three dimensions 
$$
R(\varepsilon) \cong a \left( {\varepsilon \over a} \right)^{2/5}
M^{-1/5}N^{2/5}
$$
This result provides two important observations. The first is that
the scaling $(N/M)$ is destroyed trivially by excluded volume. 
The second is that the distance constrained polymer chain
does not scale as a branched chain, when excluded volume has been taken into
account. This would mean $R \propto N^{5/(2(d+2))}$ (or in three dimensions
$R \propto N^{1/2}$), which is the Isaacson Lubensky value and 
well confirmed by experimental studies. 
Physically this means that the distance constraint is much
more restrictive and excluded volume forces are not
able to overcome this constraint. Mathematically this is also
clear, since a constraint in the partition function is much stronger
than the excluded volume (pseudo) potential.
Of course, a special regime for the constraint variable 
$\varepsilon$ exists such that the crosslinked polymer appears
as swollen branched chain. 
In this case 
$\varepsilon^2 \sim a^2M \sqrt{N}$ and 
independent of the space dimension. 

The natural question, how such types of polymers can be realized, can
be answered in such a way, that $\varepsilon$ can be viewed as the
mean distance of a "soft" crosslink, made out of a linear polymer
itself. These soft crosslinks for then themselves
a random walk of $M$ steps with a step length $a^2 \sqrt{N}$. 
This agrees with the imagination that these soft
crosslink potential change the connectivity of an original 
linear polymer. Note that the upper limit for the value
of $\varepsilon^2 \sim a^2 NM$ recovers formally the free chain. The
constraint is then physically so small that it has no
effect on the conformation.

It is clear that given size of $\varepsilon$ the number of
crosslinks cannot be arbitrary large. The largest number of 
crosslinks is of the order of
\begin{equation}
M \lesssim \left(\frac{\varepsilon}{a}\right)^2 N^{(d-2)/d}
\end{equation}
i.e., when the crosslinked polymer is given by its extremal minimal size
$R \cong a N^{1/d}$.

\subsection{Hard crosslinks $\varepsilon = 0$}

The case of hard crosslinks requires a more careful treatment
of the crosslink term. Here we use the same picture as above
where the Gaussian network has been treated. Thus the scaling
of the crosslink term is given by a random walk through the
crosslinks $R \propto M \sqrt{a^{2}(N/M)}$. 
This term is then relevant to the dominant elastic contribution in the
free energy which balances excluded volume.

For a large number of crosslinks $N>>M>>1$ the 
corresponding Flory free energy is then given by
\begin{equation}
{\cal F} = M^2 \frac{R^2}{a^2 N}+v\frac{N^2}{R^d}
\end{equation}
The size of the microgel is
\begin{equation}
R \cong a \left(\frac{N^3}{M^2}  \right)^{1/(d+2)}
\end{equation}
This result can be written in the more instructive form in three
dimensions as
\begin{equation}
R \cong a M^{1/5} \left( \frac{N}{M} \right)^{3/5}
\end{equation}
This intuitive 
equation describes a swollen microgel, where the mean strand length $(N/M)$
is fully swollen and forms a self avoiding walk. In other words
the microgel can be visualized as a $c^*$ - gel, where the blob size
defined by the strands form self avoiding walks.

In the hard crosslinked microgel the number of crosslinks cannot be 
arbitrary. As in the previous case a similar limit exists,
which would correspond to the
maximum density of the gel. The limiting density corresponds to a
densely packed network in space, i.e., to a 
size $R \cong a N^{1/d}$. Therefore this limit requires an upper bound
for the number of crosslinks
\begin{equation}
M \leq N^{(d-1)/d}  \stackrel{d=3}{=}  N^{2/3}.
\end{equation}
Clearly the latter condition holds only for dimensions less
than four, since $d=4$ is the upper critical dimension for the 
excluded volume term which is independent of the crosslink state. 

\section{Summary}

In this note we have presented reasonable scaling estimates for the size of
self crosslinked chains and microgels. Although the estimates are simple they
yield physically understandable results. The main conclusion is that the two
types of gels may exist, i.e., soft gels, where the crosslinks are extended
objects themselves, and hard gels which consists of point like
crosslinks. Their swelling behavior is completely different and fall in two
separate classes of physical systems.

\end{multicols}{2} 

\begin{references}

\bibitem{solf1} M. P. Solf and T. A. Vilgis, {\em J. Phys. A:
Math. Gen.} {\bf 28}, 6655 (1995).
\bibitem{solf2} M. P. Solf and T. A. Vilgis, {\em J. Phys I}
{\bf 6}, 1451, (1996)
\bibitem{solf3} M.P. Solf, T.A. Vilgis, {\em Phys. Rev. Lett.}
{\bf 77}, 4276, (1996)
\bibitem{kankar} Y. Kantor and M. Kardar, {\em Phys. Rev. Lett.}
{\bf 77}, 4275, (1996)
\bibitem{solf4} M.P. Solf, T.A. Vilgis, {\em Phys. Rev. E}, {\bf 55},
3037, (1997)
\bibitem{brythi} J. D. Bryngelson and D. Thirumalai,
{\em Phys. Rev. Lett.} {\bf 76}, 542 (1996).
\bibitem{branched} G. Parisi, N. Sourlas, {\em Phys. Rev. Lett.}, {\bf 46},
  871, (1981) 
\bibitem{pg} P.G. de Gennes, {\em Scaling concepts in polymer physics},
  Cornell University Press, Ithaca, 1979
\bibitem{deam} R.T. Deam, S.F. Edwards, {\em Proc. Trans. R. Soc. London}
{\bf A 260}, 317, (1976)
\bibitem{edvi} S.F. Edwards, T.A. Vilgis, {\em Rep. Progr. Phys}, {\bf 52}, 247


\bibitem{isaac} J. Isaacson, T.C. Lubensky, {\em J. Phys. Lett.(Orsay,Fr)},
{\bf 42}, 175, (1981)

\bibitem{stocky} B.  Zimm, W.H. Stockmayer, {\em J. Chem. Phys.}, {\bf 17},
  1301, (1949) 
\end{references}
\end{document}